\documentclass[10pt,aps,twocolumn,superscriptaddress,nofootinbib,tightenlines,floatfix,longbibliography]{revtex4-1}
\usepackage{amsmath,amssymb}
\usepackage{graphicx}
\usepackage{color}
\usepackage{comment}
\usepackage{bbold}
\usepackage{soul,color}
\sethlcolor{white}
\usepackage{booktabs}
\usepackage{array}
\usepackage{hyperref}
\hypersetup{
    colorlinks=true,       
    linkcolor=blue,          
    citecolor=blue,        
    filecolor=blue,      
    urlcolor=blue           
}

\def\eqref#1{{(\ref{#1})}}

\def\D{\Delta}

\def\afs{\alpha_{fs}}

\def\d{\delta}

\def\g{\gamma}
\def\s{\sigma}
\def\t{\tau}

\def\ltap{\ \raise.3ex\hbox{$<$\kern-.75em\lower1ex\hbox{$\sim$}}\ }

\def\hT{{${}^3$H}}
\def\heT{{${}^3$He}}
\def\heF{{${}^4$He}}

\begin{document}

{\count255=\time\divide\count255 by 60 \xdef\hourmin{\number\count255}
  \multiply\count255 by-60\advance\count255 by\time
  \xdef\hourmin{\hourmin:\ifnum\count255<10 0\fi\the\count255}}

\title{Quantifying the sensitivity of Big Bang Nucleosynthesis to isospin breaking \\ with input from lattice QCD}

\author{Matthew Heffernan}
\email[]{heffernan@physics.mcgill.ca}
\affiliation{Department of Physics, 
	McGill University, 3600 University Street, Montreal, QC, H3A 2T8, Canada}
\thanks{Some of this work was conducted at The College of William and Mary in Virginia by A. Walker-Loud and M. Heffernan.}

\author{Projjwal Banerjee}
\email[]{projjwal@sjtu.edu.cn}
\affiliation{Department of Astronomy,  School of Physics and Astronomy, Shanghai Jiao Tong University, Shanghai 200240, China}

\author{Andr\'{e} Walker-Loud}
\email[]{awalker-loud@lbl.gov}
\affiliation{Nuclear Science Division, 
	Lawrence Berkeley National Laboratory, Berkeley, CA 94720, USA}
\date{\today}

\begin{abstract}
We perform the first quantitative study of the sensitivity of Big Bang Nucleosynthesis to variations in isospin breaking with precise input from lattice QCD calculations.
The predicted light nuclear abundances are most sensitive to the neutron-proton mass splitting as both the initial relative abundance of neutrons to protons and the $n \rightleftharpoons p$ weak reaction rates are very sensitive to this quantity.
Lattice QCD has been used to determine this mass splitting to greater than 5-sigma, including contributions from both the down-quark up-quark mass splitting, $2\d = m_d-m_u$ and from electromagnetic coupling of the quarks to the photons with a strength governed by the fine structure constant, $\afs$.
At leading order in isospin breaking, the contribution of $\d$ and $\afs$ to $M_n-M_p$ and the nuclear reaction rates can be varied independently.
We use this knowledge and input from lattice QCD to quantitatively study variations of the predicted light nuclear abundances as $\d$ and $\afs$ are varied.
The change in the D and ${}^4$He abundances individually allow for potentially large simultaneous variations in $\d$ and $\afs$ while maintaining consistency with the observed abundances, however the combined comparison restricts variations in these sources of isospin breaking to less than $\lesssim1.25\%$ at the 3-sigma confidence level.
This sensitivity can be used to place tight constraints on prospective beyond the Standard Model theories that would modify these isospin breaking effects in the primordial Universe.
\end{abstract}

\maketitle

%
\section{Introduction} \label{intro}
%
The known baryonic mass of the Universe, which makes up only $\sim5\%$ of its energy budget, 
is composed of $\sim75\%$ H, $\sim25\%$ $^4$He and trace elements of light nuclei.
The synthesis of H and $^4$He and their relative abundance was determined in the 
first few minutes after the Big Bang in a process referred to as 
Big Bang Nucleosynthesis (BBN).  For a recent review, we refer to Ref.~\cite{Cyburt:2015mya}.
During this epoch, as the Universe expands, the temperature and density 
drops such that free nucleons fuse into mostly $^4$He and trace amounts of deuterium (D), ${}^3$He, ${}^7$Li and other light nuclei.
The remaining protons end up as H. 
A crucial quantity which impacts the relative abundance of $^4$He and H during BBN is the isospin splitting between the mass of the neutron and proton $\D M_{n-p}$.

Lattice QCD (LQCD) is a numerical, non-perturbative regulator for QCD, the fundamental theory of nuclear strong interactions, and the only rigorous tool to study the isospin splitting directly from the parameters of the Standard Model.
Recent computational advances have allowed LQCD to be used to determine the contributions to $\D M_{n-p}$ from both sources of isospin breaking.
At low energies, the two sources are the splitting between the up and down quark masses and the strength of the electromagnetic coupling between the quarks and photons.

The success of BBN in describing the observed primordial abundances of light nuclei provides strong evidence for the hot big bang model of cosmology.
In particular, the observed abundance of ${}^4$He is too high to have been produced in stars alone, but is readily produced in the BBN reactions, which assume a hot, expanding Universe.
Given the independent determination of the primordial baryon-to-photon ratio from the WMAP scan of the cosmic microwave background~\cite{Tegmark:2003ud,Spergel:2003cb,Spergel:2006hy}, and the measured light nuclear reaction rates, BBN provides a parameter free prediction of the primordial abundances of the light nuclei.
The predicted abundances match with exquisite precision the observed abundances, with the exception of ${}^7$Li, which is known as the ${}^7$Li puzzle, see for example Ref.~\cite{Cyburt:2008kw}.
Aside from this puzzle, BBN is tested against observations at the percent level, providing stringent constraints on any theories for physics beyond the Standard Model.

In this work, we provide the quantitative information necessary to constrain models which violate isospin symmetry.  For the first time, we consider the modification to light nuclear abundances predicted by BBN while simultaneously varying both Standard Model sources of isospin breaking, taking as input recent results from LQCD.
In Sec. \ref{sec:BBNreview}, we provide a brief review of Big Bang Nucleosynthesis. 
In Sec. \ref{sec:BBNiso} we explore the relation between BBN and isospin and describe the modified software we use.
We present our results in Sec.~\ref{sec:results} and then conclude.

%
\section{Review of Big Bang Nucleosynthesis \label{sec:BBNreview}}
%
Here, we provide a brief review of BBN to set the context for our work.
The most recent thorough review of BBN, which we utilized, can be found in Ref.~\cite{Cyburt:2015mya}.
Big Bang Nucleosynthesis describes the production of light nuclear elements starting 
from a nuclear statistical equilibrium of neutrons and protons and lasting for the first 
few hundred seconds of the Universe.
In Standard Big Bang Nucleosynthesis (SBBN), it is assumed that the expanding 
Universe is flat as well as isotropic and homogeneous. 
The hot expanding Universe is characterized by the photon temperature where the 
epoch of BBN starts from $T\sim 10$~MeV and lasts until $T\sim 10$~keV when neutrinos and 
photons are the only remaining relativistic species. 
During this time, the energy density of the  Universe is dominated entirely by 
relativistic species where dark matter, dark energy and ordinary 
baryonic matter make negligible 
contributions to the total energy budget. 
At temperatures $\gtrsim 2$~MeV, relativistic species, which include $e^+,e^-$, 
photons, neutrinos and anti-neutrinos are in thermal equilibrium. 
Neutrons and protons, the non-relativistic species, are kept in thermal and 
chemical equilibrium by the weak interactions
\begin{eqnarray}
n+\nu_e &\rightleftharpoons& p+e^{-},\label{eq:nprate2} \\
n+e^{+} &\rightleftharpoons& p + \bar\nu_e\label{eq:nprate3}.
\end{eqnarray}
As a result, their abundances can be calculated from nuclear statistical 
equilibrium and their relative abundance is given simply by
\begin{equation}
\label{eq:XpXn}
\frac{X_p}{X_n}=e^{(M_n - M_p)/T},
\end{equation}
where $X_n$ and $X_p$ are the neutron and proton abundances respectively, $M_n$ and 
$M_p$ are the corresponding masses and $T$ is the temperature.
As the temperature drops below $\sim 1.5$~MeV, neutrinos start to fall out of 
equilibrium as the expansion rate exceeds the weak interaction rate. 
As a result, neutrinos thermally decouple from the other species and \textit{freezeout}. 
However, weak interactions via Eqs.~(\ref{eq:nprate2}-\ref{eq:nprate3}) 
are still fast enough such that neutrons and protons remain in chemical equilibrium 
and hence Eq.~\eqref{eq:XpXn} remains valid. As the temperature drops further to $T\sim 0.9$~MeV ($\sim 1$s after the big bang), the forward reaction rates become dominant and the chemical equilibrium between neutrons and protons is no longer maintained. 
The chemical equilibrium is now maintained by 
\begin{equation}
\label{eq:npdg}
n+p\rightleftharpoons \rm{D}+\gamma\, .
\end{equation}
Using the fact that $n, p$, and D are in chemical equilibrium i.e., $\mu_d=\mu_n+\mu_p$, where $\mu_X$ is the chemical potential of species $X$, the equilibrium abundance of D can be expressed in terms of $X_n, X_p$, and
the baryon-to-photon ratio $\eta$ by the relation 
\begin{equation}
X_{\rm D}= 5.66\times10^{-4}\eta \left(\frac{T}{{\rm MeV}}\right)^{3/2} \exp\left(\frac{B_{\rm D}}{T}\right)X_nX_p\, ,
\end{equation}
where the deuterium binding energy is
\begin{equation}
B_{\rm D} \simeq 2.224 \textrm{ MeV}\, .
\end{equation}
The prefactor of 5.66 comes from the relation
\begin{equation*}
\frac{2^4}{\sqrt{\pi}} \zeta(2)\frac{g_d g_\gamma}{g_p g_n}
	\left(\frac{m_p}{\rm MeV}\right)^{-3/2} \simeq 5.66\, ,
\end{equation*}
where $g_i$ is the number of internal degrees of freedom of particle $i$:
$g_p=g_n=g_\gamma=2$ and $g_d=3$.
The proton mass is $m_p=938.27$~MeV and $\zeta(2)\approx 1.202$.
Although the rate 
for Eq.~\eqref{eq:npdg} is much faster than the expansion rate, the equilibrium abundance of 
D is very low as there are many photons with energy higher than $B_{\rm d}$.
As a result, the rate of production of heavier isotopes from D,  
which depend on the D abundance, are extremely low. Hence, all isotopes heavier than 
D are chemically decoupled and the only species which are in chemical equilibrium are neutrons, protons and D.

This delay in the production of heavier isotope is called the \textit{deuterium bottleneck}. 
As the temperature drops below $\sim 0.08$~MeV ($\sim 200$~s), the number of photons with 
energy $\gtrsim 2.224$ MeV relative to the D abundance becomes less than unity. This allows the 
equilibrium  D abundance to become high enough for heavier isotopes to form 
and marks the end of the deuterium bottleneck. Although, the duration of the  
of this bottleneck is much shorter than the neutron decay time, 
the weak interactions convert much of neutrons 
into protons through the reactions Eq.~\eqref{eq:nprate2} and Eq.~\eqref{eq:nprate3} during the first $\sim 30$ s followed by the usual neutron beta decay 
\begin{equation}
n\rightarrow p +e^{-}+\bar\nu_e\, ,
\end{equation}
until the end of the deuterium bottleneck.
Since the binding energy of $^4$He, $B_{^4\rm{He}} = 28.3$~MeV, is much higher than any neighboring isotopes, which have binding energies $\lesssim 8$ MeV, it is by far the most abundant 
isotope populated as the system tries to achieve chemical equilibrium.   
As as a result, the majority of free neutrons available at this time combine with protons to form 
$^4$He by $\sim 400$~s while other nuclei such as D, 
$^3$H, $^3$He, $^6$Li, and $^7$Li make up $\lesssim 10^{-4}$ of the total 
mass fraction of the baryonic matter of the Universe. 
The remaining free protons, which constitute $\sim75$\% by mass fraction, end up as $^1$H. 
The abundance of lighter elements continues to evolve up to $\sim 300 $ minutes after the Big Bang, 
at which point the temperature $T\lesssim 10$~KeV, becomes too low for fusion reactions to occur. 
The detailed nucleosynthesis involves a complex chain of reactions in which the isotopes 
are out of equilibrium but can be calculated precisely with a nuclear reaction network.
These processes were first worked out in Refs.~\cite{PhysRev.73.803,PhysRev.74.1737,Wagoner:1966pv}.  We utilize a code based on that developed by Kawano~\cite{Kawano:1988vh,Kawano:1992ua,Smith:1992yy}, which serves as the basis for many modern BBN codes.

\section{BBN and isospin \label{sec:BBNiso}} 
\noindent

There have been a number of articles considering the sensitivity of BBN to variations 
in the fundamental parameters of the Standard Model, including the electromagnetic fine structure constant~\cite{Kolb:1985sj,Campbell:1994bf,Bergstrom:1999wm,Nollett:2002da,Chamoun:2005xr},
the quark masses~\cite{Calmet:2001nu,Dent:2001ga,Flambaum:2002de,Beane:2002vs,Dmitriev:2002kv,Flambaum:2002wq,Kneller:2003xf,Dmitriev:2003qq,Muller:2004gu,Landau:2004rj,Chamoun:2005xr,Coc:2006sx,Flambaum:2007mj,Berengut:2009js,Bedaque:2010hr,Berengut:2013nh,Epelbaum:2013wla}, 
and recently, a more general 
consideration of the naturalness of the weak scale in a multiverse context~\cite{Hall:2014dfa}.
Our understanding of the quark mass dependence of seemingly finely tuned energy levels has been extended to the Hoyle State~\cite{Epelbaum:2011md} and the sensitivity of the production of ${}^{12}$C to variations in the average light quark mass~\cite{Epelbaum:2012iu}.

The recent studies involving the quark mass variation have considered the sensitivity of BBN to varying the average light quark mass $\hat{m} = \frac{1}{2}(m_u+m_d)$ with input from lattice QCD~\cite{Bedaque:2010hr,Epelbaum:2012iu,Berengut:2013nh,Epelbaum:2013wla}.
Previous works that considered sensitivity to the fine structure constant relied upon an old estimate of the electromagnetic self-energy contribution to $M_n-M_p$.
If electromagnetic effects were the only contribution to the nucleon mass splitting, the proton would be heavier than the neutron, the opposite of what is observed in nature.
The second source of isospin breaking is therefore at least as important as that from QED but it has previously not been considered in studies of BBN sensitivity to parameters of the SM.

The two sources of isospin breaking within the SM relevant to BBN are the electromagnetic fine structure constant, $\afs$, and the splitting between the down and up quark masses, 
\begin{equation}
	\d = \frac{1}{2}(m_d - m_u)\, .
\end{equation}
As the quark masses are generated by their interaction with the Higgs field, we are really considering the variation of the dimensionless up and down quark Yukawa couplings to the Higgs field while holding the Higgs vacuum expectation value fixed with respect to the scale of strong interactions, $\Lambda_{QCD}$.
Further, we are considering variations of $\d$ while holding fixed the average value of the light quark mass $\hat{m}$, or rather, the average value of the Yukawa couplings.

In this work, we restrict our attention to the production of D and ${}^4$He in BBN to constrain the possible variations of $\d$ and $\afs$.
The dominant sensitivity of BBN to isospin arises from changes in the initial abundances of protons and neutrons via Eq.~\eqref{eq:XpXn}, through variations in the neutron-proton mass splitting
\begin{equation}\label{eq:mn-mp}
\D M_{n-p} = M_n - M_p\, ,
\end{equation}
and through modifications of the weak $n\rightleftharpoons p$ rates, described in more detail in Sec.~\ref{sec:nprates}.
These reactions can be parameterized as depending upon the nucleon mass splitting, Eq.~\eqref{eq:mn-mp} as well as sub-dominant radiative electromagnetic corrections which are sensitive to $\afs$ at the sub-percent level, see Refs.~\cite{Dicus:1982bz,Lopez,Esposito,Smith}.
The final abundances of H and ${}^4$He are set almost entirely from these two effects.

The binding energy of D plays an important role since it determines the duration of the deuterium bottleneck 
during which neutrons are converted into protons. Consequently, it directly impacts the abundance of H and {\heF} with a higher value of $B_{\rm D}$ resulting in higher {\heF} and lower H and vice-versa.
However, to leading order in isospin breaking, we do not have to consider modifications of $B_{\rm D}$ due to isospin violation as the deuteron is an isoscalar particle, and therefore modifications to the binding energy would begin at second order in isospin breaking.
The abundance of D is very sensitive to $\D M_{n-p}$ and the $n\leftrightharpoons p$ rates as these impact the abundance of free neutrons to form D.
In addition, D abundance is sensitive to $\afs$ via the Coulomb repulsion in charged light particle fusion reactions involving D and indirectly via  $np\leftrightharpoons \rm{D}\g$ rates which depend on the electromagnetic coupling.
Similarly, the production of {\heT} and tritium, \hT, are also sensitive to $\afs$.
In principle, we should consider the relative modification to the binding energies of {\heT} and {\hT} as they form and isodoublet.
However, whichever nucleus is heavier will decay to the other in a lifetime sensitive to the mass splitting between them.
As both {\heT} and {\hT} are pathways to the formation of {\heF}, we do not try and estimate the modification to the binding energies of these $A=3$ nuclei, as the final abundance of {\heF} is not sensitive to these variations.
Instead, we convert all $A=3$ nuclei to {\heT} after BBN stops as a proxy for the stable nucleus.

\subsection{$n\leftrightharpoons p$ rates \label{sec:nprates}}

At zero temperature, the free neutron decay rate (inverse lifetime) is given by 
\begin{equation}
\Gamma_{n\rightarrow p} = \frac{(G_F \cos \theta_C)^2}{2\pi^3} m_e^5\, \frac{g_V^2+3g_A^2}{4}
	f\left( \frac{\D M_{n-p}}{m_e} \right),
\end{equation}
where $G_F$ is the weak Fermi-constant, $\theta_C$ is the weak Cabibbo angle, $m_e$ is the electron mass, $g_V=1$ is the nucleon vector coupling and $g_A=1.2723$ is the nucleon axial vector coupling.
The quantity $\frac{1}{4}(g_V^2+3g_A^2) f(z)$ represents an integral over the phase space of the decay and depends only upon the nucleon isovector mass normalized by the electron mass.
If the nucleons were point particles, the phase space integral could be determined analytically, as for example in Ref.~\cite{Griffiths:2008zz},
\begin{align}\label{eq:f_point}
f_{pt}(z) &= \frac{1}{15} (2z^4 - 9z^2 - 8) \sqrt{z^2 -1}
\nonumber\\&\phantom{=}
	+z \ln \left( z + \sqrt{z^2 -1} \right)\, .
\end{align}
Using this point expression, the neutron lifetime is calculated to be $\t_n^{pt} \simeq 946$~s, which is in the right ballpark of the experimentally measured lifetime, $\t_n^{PDG} = 880.2\pm 1.0$~s~\cite{Olive:2016xmw,Mampe:1993an,Byrne:1996zz,Serebrov:2004zf,Pichlmaier:2010zz,Steyerl:2012zz,Yue:2013qrc,Arzumanov:2015tea}.
An accurate theoretical determination of $\frac{1}{4}(g_V^2+3g_A^2)f(z)$ is not yet possible as this depends upon detailed structure of the nucleon, which is generated by QCD.
Only very recently has LQCD been successfully used to compute the nucleon axial coupling $g_A$ directly from QCD~\cite{Berkowitz:2017gql}.
In the next few years, a full calculation of the neutron decay amplitude should be possible directly with LQCD.

In the standard BBN codes, these integrals are performed using this point approximation, including radiative QED corrections and finite temperature modifications, and absolutely normalized to reproduce the measured neutron lifetime at zero temperature.
Using the formula with radiative QED corrections ($\t_n^{pt,QED} \simeq 944.1$~s), this leads to an overall normalization for the lifetime and  $n\leftrightharpoons p$ reactions of $880.2 / 944.1$.

While we are unable to determine the decay amplitude directly, we can make the following reasonable assumption:
as we vary the isospin breaking corrections and hence $\D M_{n-p}$, the relative change in the point approximation, Eq.~\eqref{eq:f_point} provides a good estimate for the change full decay probability.
This assumption relies on the shape of the integrand, which is dictated by the internal nucleon structure arising from QCD being insensitive to isospin breaking.
The modified decay rate at zero temperature is then given by
\begin{equation}
\Gamma_{n\rightarrow p}(\d,\alpha_{f.s.}) = \frac{1}{\t_n^{PDG}}
	\frac{f_{pt} \Big(\D M_{n-p}(\d,\alpha_{f.s.})\Big)}
	{f_{pt}\Big(\D M_{n-p}(\d^{phys},\alpha_{f.s.}^{phys})\Big)},
\end{equation}
and similarly for the other $n\leftrightharpoons p$ reactions at zero and finite temperature.
We emphasize that a similar assumption is made to compute the various $n\leftrightharpoons p$ reaction rates at finite temperature (with the physical values of $\d^{phys}$ and $\alpha_{f.s.}^{phys}$) in the standard BBN computations.  
This limitation arises from our current inability to directly compute these rates, or the required nucleon structure with sufficient precision, directly from QCD.

\subsection{The neutron-proton mass splitting}
The dominant sensitivity of BBN to isospin breaking arises from dependence upon the 
neutron proton mass splitting, which is known very precisely~\cite{Mohr:2008fa,Nakamura:2010zzi,Codata:2012}
\begin{equation}
\label{eq:pdgMn-Mp}
\D M_{n-p}^\textrm{PDG} \equiv M_n - M_p = 1.29333217(42) \textrm{ MeV}\, .
\end{equation}
At LO in isospin breaking, we can separate this isovector mass splitting into two contributions
\begin{equation}
	\D M_{n-p} = \D M^\g_{n-p} + \D M^{\d}_{n-p} + \cdots\, ,
\end{equation}
which arise from QED, $\D M^\g_{n-p}$ and QCD, $\D M^{\d}_{n-p}$ respectively.%
\footnote{There is also a contribution from $Z$ boson couplings but it is significantly 
smaller than we will be concerned with.} 
Both the electromagnetic and strong contributions scale linearly in the corresponding isospin breaking parameters, $\D M^\g_{n-p} \propto \afs$ and $\D M^{\d}_{n-p} \propto \d$.
Beyond LO, $\cdots$, these contributions receive small quadratic corrections and also become entangled as the quark mass operators are needed to 
renormalize ultraviolet divergences in the QED self-energy~\cite{Gasser:1974wd,Collins:1978hi}

Understanding the neutron-proton mass splitting from first principles has been an intriguing 
topic for many decades, see for example Ref.~\cite{Zee:1971df}.
Before we knew of quarks and gluons, it was proposed that the magnetic structure of the nucleon 
could give rise to the excess mass of the neutron over the proton~\cite{Feynman:1954xxl}.
This idea was later fully developed using dispersion theory~\cite{Cini:1959} and is now know as 
the Cottingham formula~\cite{Cottingham:1963zz}:
For a general hadron, the EM self-energy can be determined through its forward Compton scattering 
amplitude by use of dispersion relations.
For isospin 0 and 1 mass combinations, such as the neutron-proton mass splitting, one must use a 
subtracted dispersion integral, as the Compton amplitude otherwise leads to a non-vanishing 
contribution on the infinite arc in the Cauchy Integral~\cite{Harari:1966mu,Abarbanel:1967zza}.
The need for a subtraction leads to an unfortunately large uncertainty~\cite{Gasser:1974wd,WalkerLoud:2012bg,Thomas:2014dxa,Erben:2014hza} 
with the original estimate~\cite{Gasser:1974wd} of $\D M^\g_{n-p} = -0.76 \pm 0.30$~MeV and an 
updated determination~\cite{WalkerLoud:2012bg} with modern knowledge of the structure of the 
nucleon resulting in $\D M^\g_{n-p} = -1.30 \pm 0.47$~MeV.%
\footnote{Ref.~\cite{Gasser:2015dwa} criticized Ref.~\cite{WalkerLoud:2012bg} and suggested the original estimate of Ref.~\cite{Gasser:1974wd} stands, but did not provide an updated value or uncertainty.  However, the central value of Ref.~\cite{Gasser:1974wd} is in striking disagreement with $\D M_{n-p}^\textrm{PDG} - \D M_{n-p}^\d$ as determined from LQCD: $\D M_{n-p}^\textrm{PDG} - \D M_{n-p}^\g\text{\cite{Gasser:1974wd}} = 2.05$~MeV versus $\D M_{n-p}^\d[\textrm{LQCD}] = 2.39(12)$~MeV.  
Ref.~\cite{Gasser:1974wd} is also inconsistent with the LQCD values of $\D M_{n-p}^\g$.  Both of these LQCD results are discussed below.
Therefore, either there are beyond the Standard Model contributions to $\D M_{n-p}$ or the result of Ref.~\cite{Gasser:1974wd} is incorrect.}
Previous studies of the sensitivity of BBN to variations in $\afs$ relied upon this old determination of $\D M^\g_{n-p}$.

Lattice QCD provides a rigorous means of determining both the QCD and QED contributions to $\D M_{n-p}$.
Including QED effects in LQCD calculations introduces a new IR (infrared) systematic which must be controlled.
This new IR systematic is induced by compressing the photons to fit in the small periodic boxes 
typically used in calculations, with sizes $\sim2-4$~fm per side.
The calculations of $\D M^\g_{n-p}$ are mostly still preliminary~\cite{Blum:2010ym,Horsley:2015eaa} 
with only two calculations performed which have addressed all the systematics~\cite{Borsanyi:2013lga,Borsanyi:2014jba} and only the latter one, controlling them~\cite{Borsanyi:2014jba}.
The first calculation determined $\D M^\g_{n-p}\text{\cite{Blum:2010ym}} = -0.38\pm0.07$~MeV 
(statistical uncertainty only), while the calculations addressing all systematics found 
$\D M^\g_{n-p}\text{\cite{Borsanyi:2013lga}} = -1.59\pm0.46$~MeV and 
$\D M^\g_{n-p}\text{\cite{Borsanyi:2014jba}} = -1.00\pm0.16$~MeV respectively.
There are a number of new ideas introduced recently to improve the inclusion of QED with LQCD 
calculations~\cite{Lucini:2015hfa,Lehner:2015bga,Endres:2015gda} including the use of a finite 
photon mass~\cite{Endres:2015gda}.  While these ideas are very promising, they have not yet 
been used yet to make final predictions of the QED contribution to $\D M_{n-p}^\g$.

\begin{figure}
\includegraphics[width=0.99\columnwidth]{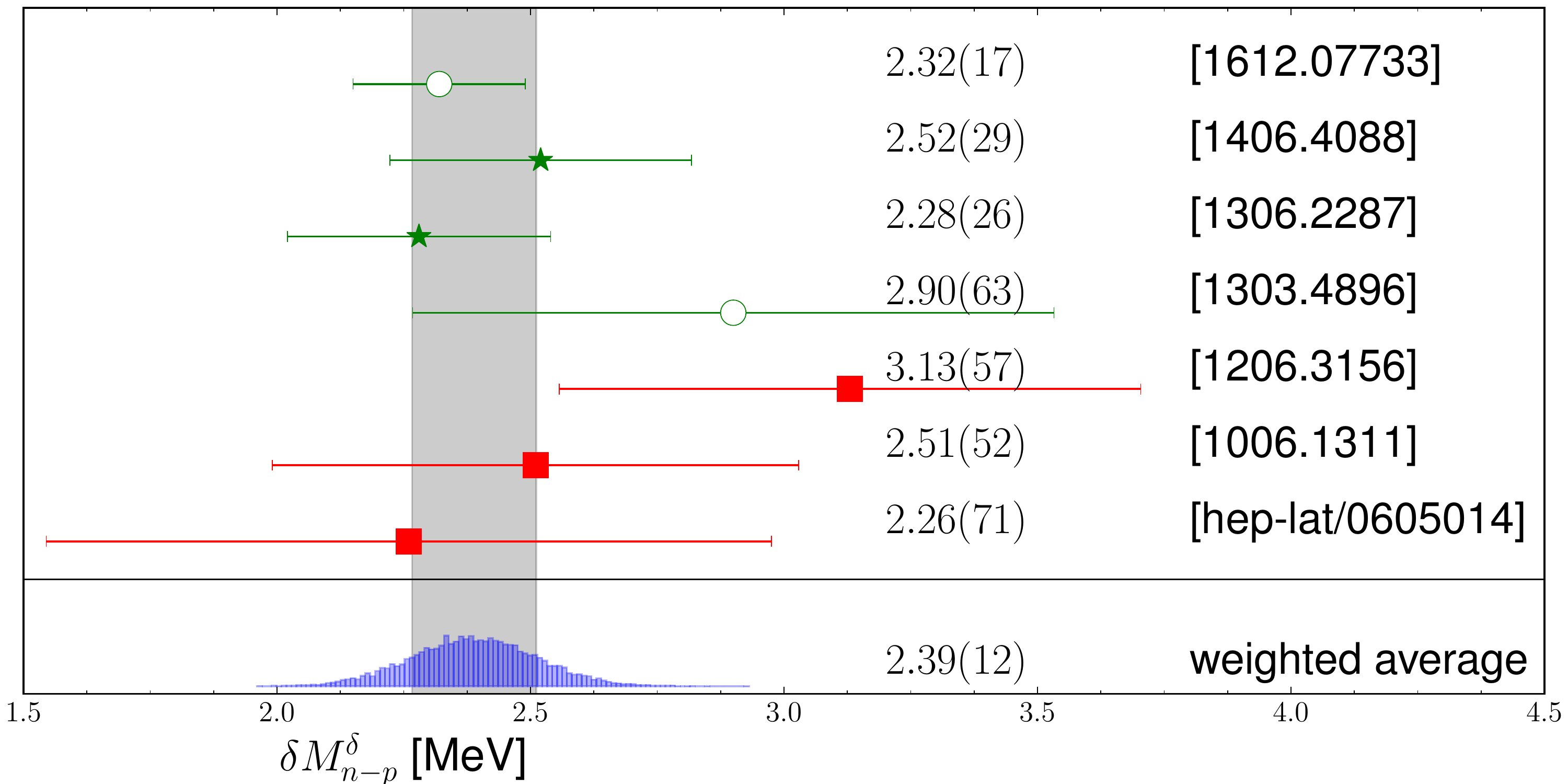}
\caption{\label{fig:dmN_mdu}Average of lattice QCD calculations of $\D M_{n-p}^{\d}$ as described in the text.  }
\end{figure}

Alternatively, we can determine the electromagnetic self-energy contribution by comparing the 
experimental splitting with the LQCD determination of the QCD contribution, $\D M_{n-p}^\d$ while 
assuming there are no beyond the Standard Model contributions to the neutron-proton mass splitting.
In contrast to the QED calculations, there are a number of LQCD calculations of the down-quark up-quark mass splitting 
contribution to $\D M_{n-p}^\d$~\cite{Beane:2006fk,WalkerLoud:2009nf,Blum:2010ym,deDivitiis:2011eh,Horsley:2012fw,Shanahan:2012wa,deDivitiis:2013xla,Borsanyi:2013lga,Borsanyi:2014jba,Brantley:2016our}.
In Fig.~\ref{fig:dmN_mdu}, we show most of the LQCD calculations of $\D M_{n-p}^\d$, along with a weighted average, as described below.
There is an additional lattice QCD calculation of $\D M_{n-p}$ including both strong isospin breaking and QED corrections~\cite{Horsley:2015eaa}, but the separation into the QED and QCD parts was performed in a different scheme, the Dashen scheme, so a direct comparison is not possible without a conversion.
We use a FLAG-inspired color scheme~\cite{Aoki:2016frl} for the various results, where a red square indicates at least 
one systematic has not been fully addressed, such as a continuum limit.  
A solid green star indicates all systematics are controlled while an open green circle indicates the systematics are somewhere between these two options.
There are only two results with fully controlled systematics~\cite{Borsanyi:2013lga,Borsanyi:2014jba}, so we choose to include the others in a weighted average, similar to that in Ref.~\cite{Junnarkar:2013ac}.
We chose to weight each result with $w_i = y_i / \sigma_i^2$ where $\sigma_i$ is the total uncertainty 
added in quadrature from the $i^{th}$ result and $y_i=3$ for the green star results, $y_i=2$ for the 
open green circle results and $y_i=1$ for the red square results.
This results in
\begin{equation}
\D M_{n-p}^\d = 2.39(12) \textrm{ MeV}\, ,
\end{equation}
also depicted in Fig.~\ref{fig:dmN_mdu} as the central gray band.
This weighting procedure is sampled $10^4$ times to produce the weighted distribution shown at the bottom of the figure.
As mentioned above, this also allows us to estimate the QED contribution using the experimental value
\begin{equation}
\D M_{n-p}^\g = \D M_{n-p}^\textrm{PDG} - \D M_{n-p}^\d = -1.10(12) \textrm{ MeV}\, ,
\end{equation}
in very good agreement with the direct determination of Ref.~\cite{Borsanyi:2014jba}.

With these two corrections determined directly and indirectly from LQCD, we can parameterize the variation of $\D M_{n-p}$ in terms 
of the fundamental isospin breaking parameters through the relations
\begin{align}\label{eq:dmN_d}
\D M_{n-p}^\d &= \phantom{+} 2.39\, \frac{\d}{\d^{phys}} \textrm{ MeV}\, ,
\\
\label{eq:dmN_g}
\D M_{n-p}^\g &= -1.10\, \frac{\alpha_{f.s.}}{\alpha_{f.s.}^{phys}} \textrm{ MeV}\, .
\end{align}
This allows us to precisely and accurately track our first principles understanding of $\D M_{n-p}$ from LQCD through the BBN reactions.
Specifically,  we explore the effect of variation of $\d$ and $\afs$ from the currently measured values at the time of BBN such that production of light elements is still consistent with the currently measured values. We restrict our attention to the abundances D and ${}^4$He in 
order to constrain  $\d$ and $\afs$. 

\subsection{BBN Software}
The detailed nucleosynthesis was followed using a publicly available BBN code (see \cite{Sarkar_url}). 
The code includes a reaction network of 26 isotopes from $^1$H to $^{16}$O and 91 reactions connecting the various isotopes. 
The code was updated to include latest reactions rates and isotope masses. 
We incorporate the zero-temperature radiative QED corrections to the  $np\leftrightharpoons \rm{D}\g$ rates as prescribed in Refs.~\cite{Dicus:1982bz,Lopez,Esposito,Smith}.
Other corrections to the $n\rightleftharpoons p$ rate due to finite nucleon mass, finite temperature radiative and QED corrections,  as well as corrections due to incomplete neutrino decoupling was not included. 
These effects are discussed in \cite{Lopez} and amount to $\sim 0.72\%$ increase in the final $^4$He abundance with a corresponding downward change in H. 
The final $^4$He and H abundance was adjusted accordingly where the dependence of these corrections on $\D M_{n-p}$ was neglected. 
The value of the effective weak interaction matrix element, which normalizes the $n\rightleftharpoons p$ rates, was calculated such that for zero temperature the neutron life time reduced to the current recommended measured value of 880.2~s~\cite{Olive:2016xmw} when the currently measured value in Eq.~\eqref{eq:pdgMn-Mp} was used.

To facilitate a user friendly method of running these calculations, we additionally adopted the Fortran code with a simple Python interface. 
The code was modified with string placeholders for values we wished to modify.  For a given pair of $\afs$ and $\d$ values, we computed the resulting value of $\D M_{n-p}$ using Eqs.~\eqref{eq:dmN_d} and \eqref{eq:dmN_g}, and then replaced the values of $\D M_{n-p}$ and $\afs$ in the code, compiled on the fly, ran the BBN calculation, and then collected the output in an HDF5 file using the PyTables interface.  
In order to diagnose and determine the dominant contributions to the change in various mass fractions, we allowed for an independent variation of the value of $\afs$ that entered the weak and thermal reaction rates.
Our modified code and Python interface is made available to interested parties with this article.

\section{Results \label{sec:results}}

Using the software described above, we ran the SBBN code for independent variations of $\afs$ and $\d$ to determine the resulting primordial abundances of H, D, and ${}^4$He.
The standard BBN parameters of 3 generation of neutrino species with zero chemical potential was used. 
The value of baryon-to-photon ratio $\eta$ was taken to be $\eta= (6.05\pm0.07)\times10^{-10}$ \cite{Olive:2016xmw} which is consistent with the latest \textit{Planck} measurements of the Cosmic Microwave Background~\cite{Ade:2013zuv}.
Our modified code was calibrated by comparing the computed primordial abundances with their recent best estimates~\cite{Aver:2013,Aver:2015iza} for the values of $\afs=\afs^{phys}$ and $\d=\d^{phys}$.

\begin{figure}
\includegraphics[width=\columnwidth]{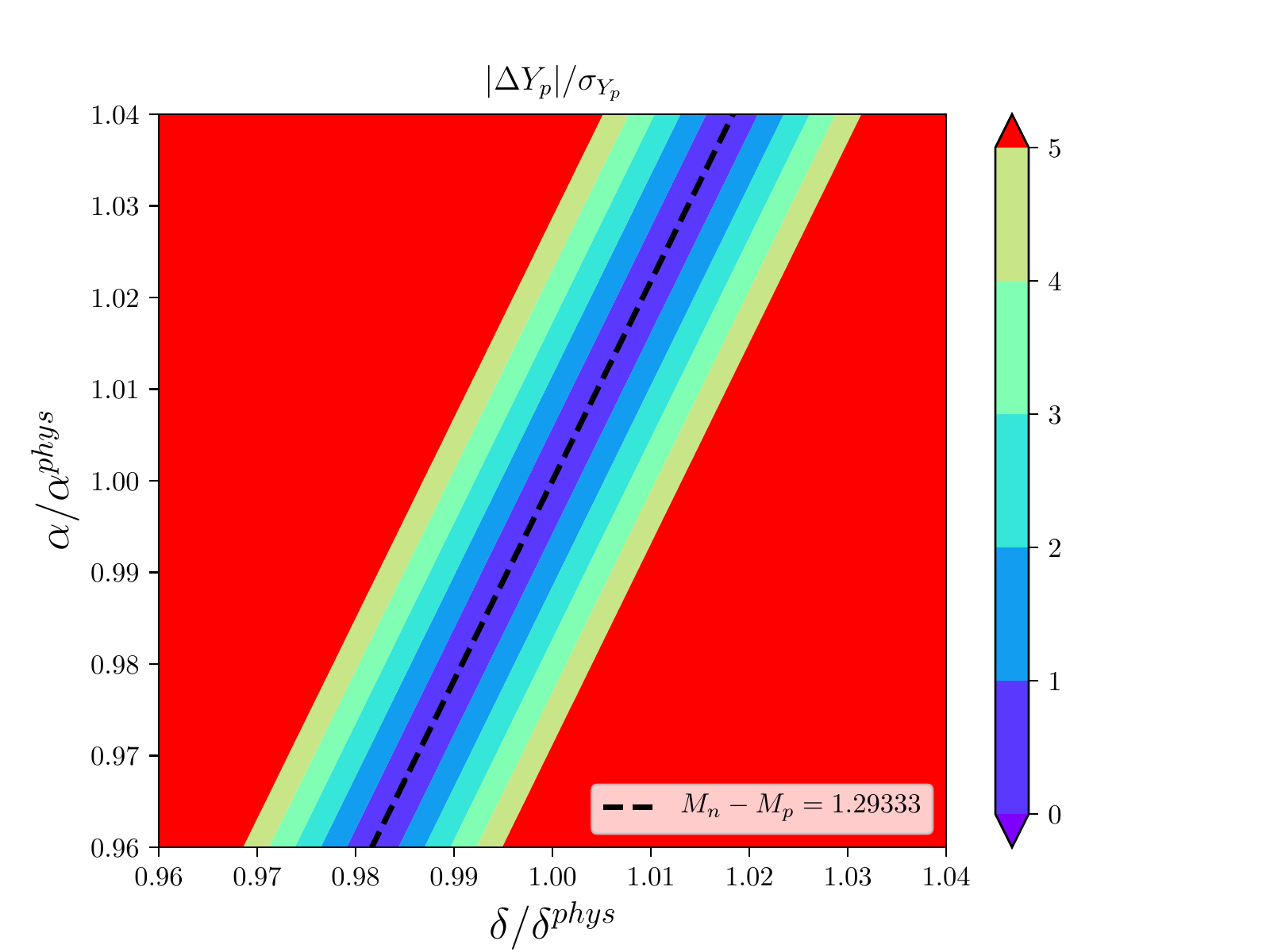}
\caption{\label{fig:he4_contour} Contour plot of the $n$-sigma change in the ${}^4$He mass fraction as a function of $\afs/\afs^{phys}$ and $\d/\d^{phys}$ measured with respect to the observed uncertainty in $Y_p$, $\s_{Y_p}$.  The dashed line is the line of constant $\D M_{n-p} = \D M_{n-p}^{phys}$.}
\end{figure}

In Fig.~\ref{fig:he4_contour}, we plot the quantity
\begin{equation*}
z_\textrm{He} 
	= \frac{|\D Y_p|}{\s_{Y_p}}
	= \frac{ |Y_p(\afs,\d) - Y_p(\afs^{phys},\d^{phys})|}{\s_{Y_p}^{obs.}}
\end{equation*}
where $Y_p(\afs,\d)$ is the computed ${}^4$He mass fraction as a function of $\afs$ and $\d$ and $\s_{Y_p}^{obs.}$ is the best estimate for the observed uncertainty of the primordial ${}^4$He mass fraction~\cite{Aver:2013,Aver:2015iza}.  
Each contour represents an $n$-sigma change in the mass fraction as compared to the mass fraction computed at $\afs=\afs^{phys}$ and $\d=\d^{phys}$, normalized by $\s_{Y_p}^{obs.}$.
To guide the eye, we also plot the line of constant $\D M_{n-p} = \D M_{n-p}^{phys}$, which is nearly parallel with the line of constant $z_\textrm{He}$.

\begin{figure}
\includegraphics[width=\columnwidth]{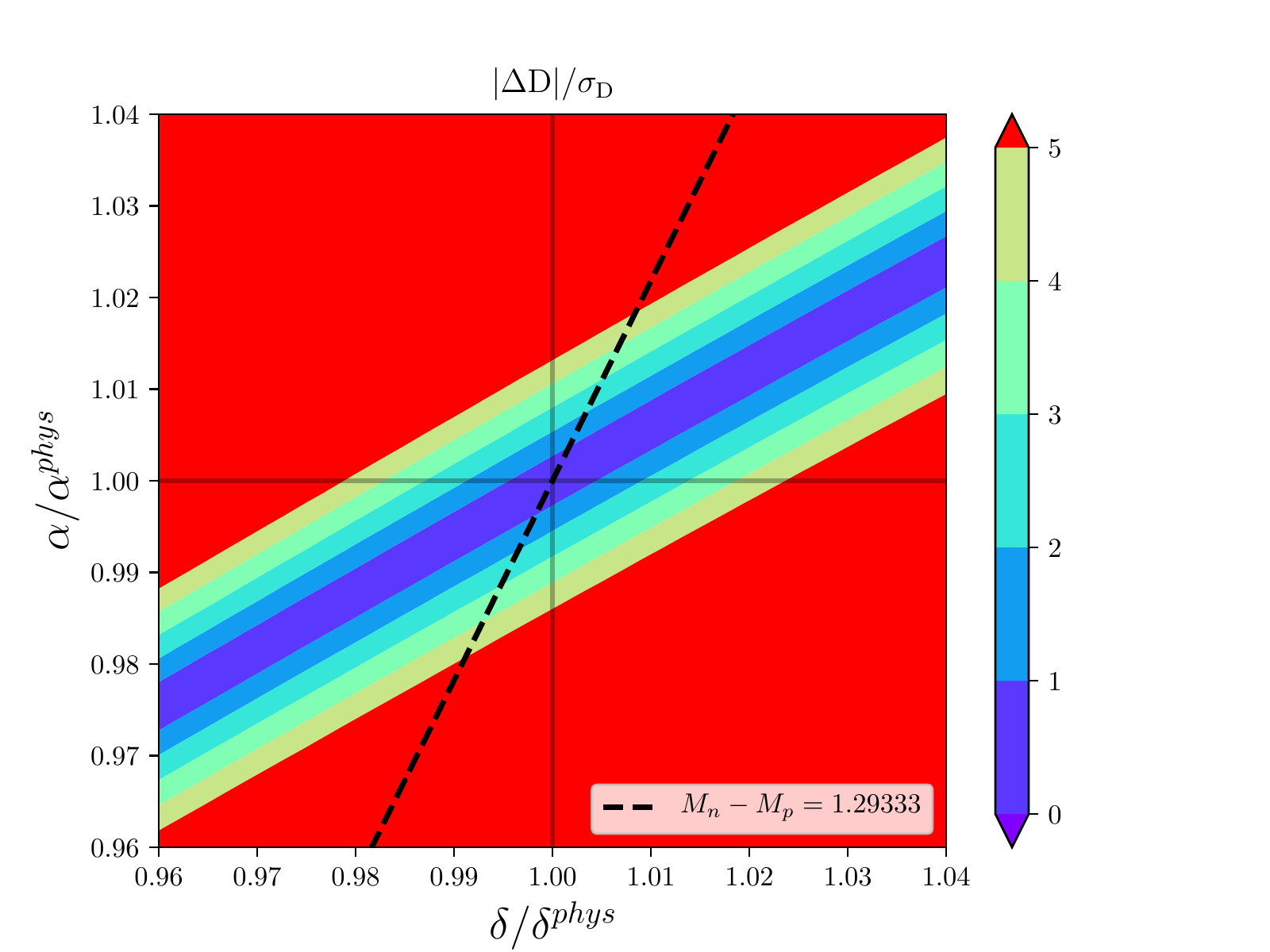}
\caption{\label{fig:d_contour} Same as Fig.~\ref{fig:he4_contour} for the $n$-sigma change in the D number fraction normalized with respect to H.}
\end{figure}

In Fig.~\ref{fig:d_contour}, we similarly plot the computed change in the deuterium abundance normalized with the best estimate of the observed uncertainty, $\s_{\rm{D}/\rm{H}}$,
\begin{equation*}
	z_{\rm D} = \frac{|\D \rm{D}/\rm{H}|}{\s_{\rm{D}/\rm{H}}} = \frac{| \rm{D}/\rm{H}(\afs,\d) - \rm{D}/\rm{H}(\afs^{phys},\d^{phys})|}{\s_{\rm{D}/\rm{H}}}
\end{equation*}
We see clearly that both $\afs$ and $\d$ can potentially be varied by $\gtrsim 10\%$ from their currently measured values without affecting the abundances of ${}^4$He and D individually, provided the changes are correlated to lie along lines of constant $z$ for each isotope.
As is clear from these figures, the slopes of these lines are not parallel, and so a combined analysis will yield a constraint on the allowed variation of $\afs$ and $\d$.

\begin{figure}
\includegraphics[width=0.99\columnwidth]{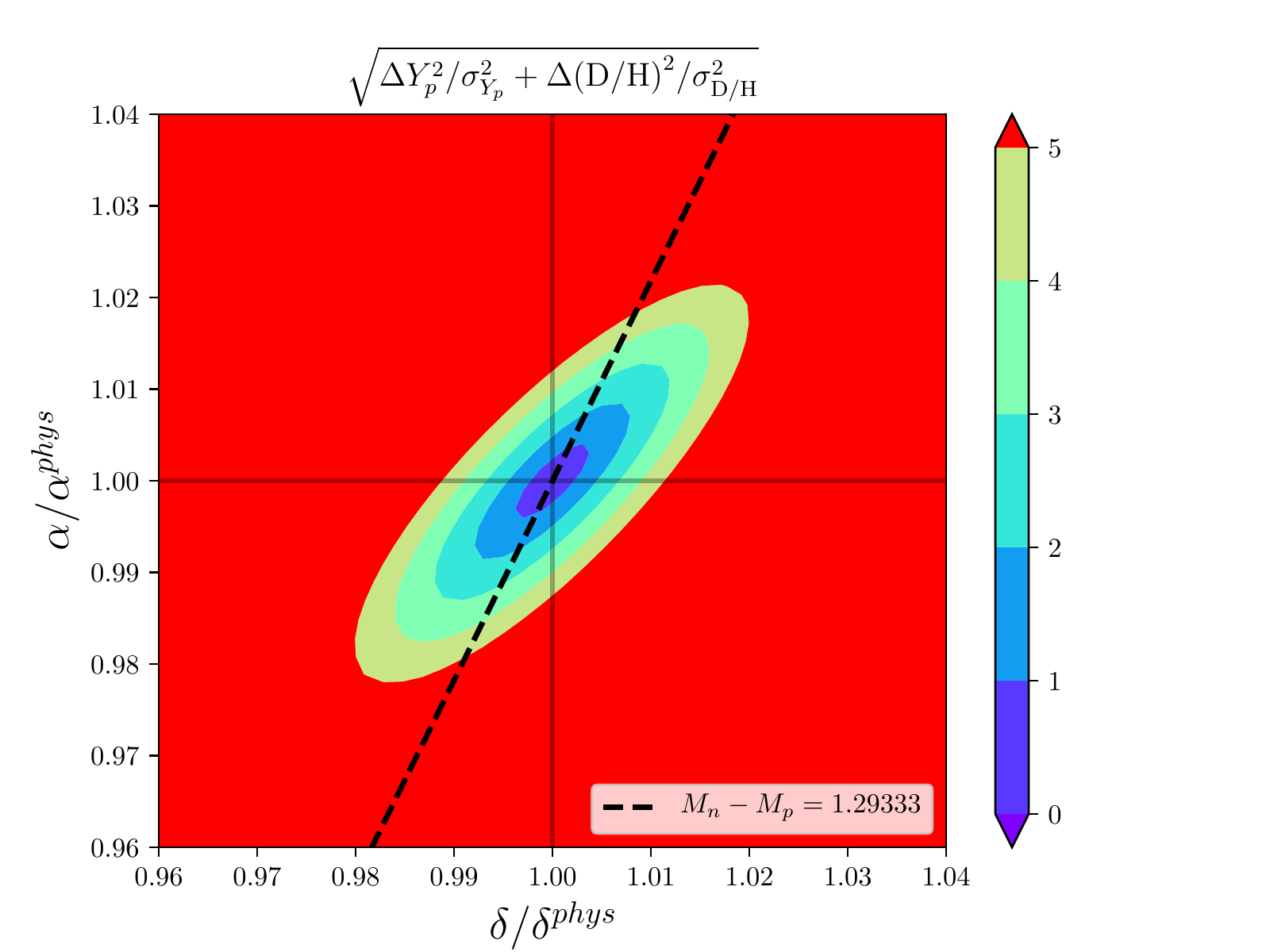}
\caption{\label{fig:X_contour}  Same as Fig.~\ref{fig:he4_contour} for the combined $n$-sigma change in the D and $^4$He.}
\end{figure}

In Fig.~\ref{fig:X_contour}, we plot the $n$-sigma variation of $z_X$
\begin{equation*}
z_X^2 = 
	\frac{(\D Y_p)^2}{(\s_{Y_p})^2} + \frac{(\D \rm{D}/\rm{H})^2}{(\s_{\rm{D}/\rm{H}})^2}\, .
\end{equation*}
At the 3-sigma level, the combined analysis constrains the variation of $\afs$ and $\d$ to be $\lesssim \pm 0.8\%$ if they are varied independently.  
If $\afs$ and $\d$ are varied together in a correlated manner, they variations are constrained to be $\lesssim \pm 1.25\%$.

\begin{figure}
\includegraphics[width=\columnwidth]{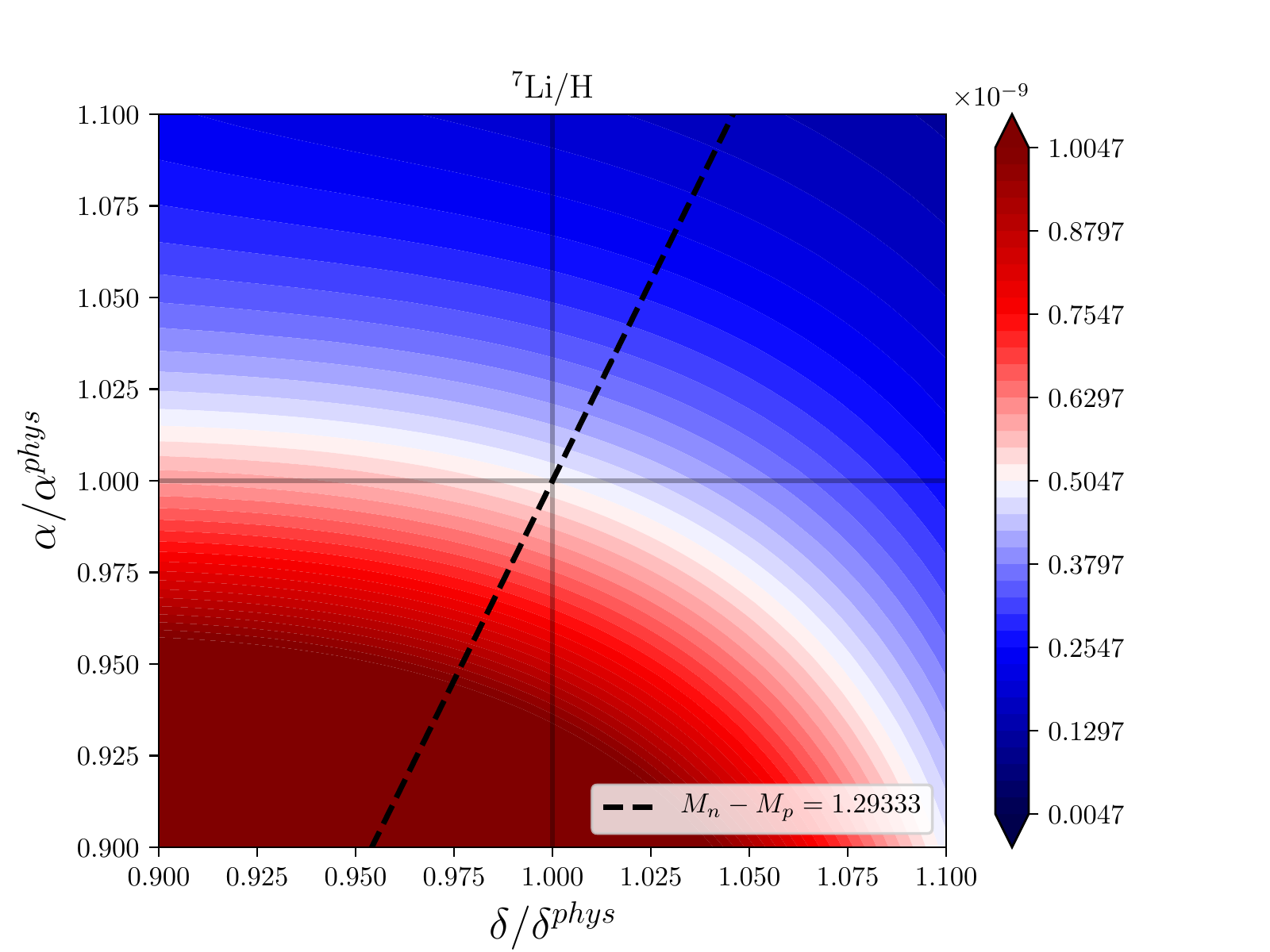}
\caption{\label{fig:Li7_density} Contour plot of the predicted ${}^7$Li/H number fraction as a function of $\afs$ and $\d$.}
\end{figure}

Finally, we rule out the possibility that primordial variations in $\afs$ and $\d$ could resolve the ${}^7$Li puzzle.
In Fig.~\ref{fig:Li7_density}, we plot the predicted number fraction of $^7$Li.%
\footnote{The lifetime of $^7$Be is longer than the time scale of BBN, so all final $^7$Be is converted to $^7$Li to determine the predicted $^7$Li abundance.} 
In order to reduce $^{7}$Li/H to the upper limit of the measured value of $(1.6 \pm 0.3) \times 10^{-10}$ \cite{Sbordone}, the values of $\afs$ and $\d$ lie in a region of greater than 5-sigma discrepancy from the combined D and $^4$He constraints.

\begin{figure}
\includegraphics[width=\columnwidth]{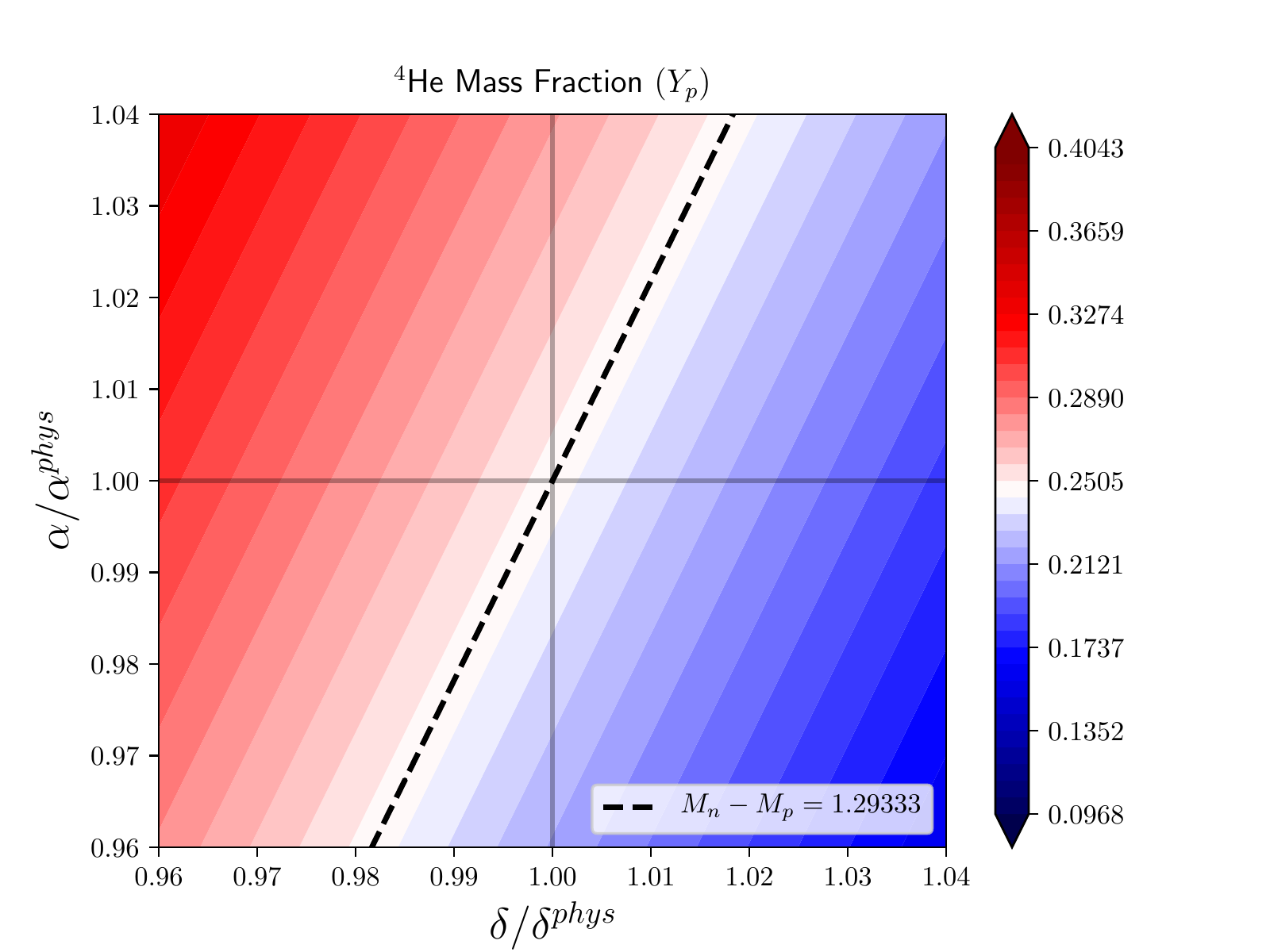}
\caption{\label{fig:he4_density} Contour plot of the absolute ${}^4$He mass fraction as a function of $\afs$ and $\d$.}
\end{figure}

As a final academic exercise, we plot the absolute ${}^4$He mass fraction as a function of $\afs$ and $\d$ in Fig.~\ref{fig:he4_density}.  As noted above, the ${}^4$He mass fraction tracks very closely the nucleon mass splitting, $\D M_{n-p}$.  Very small changes in either of the fundamental isospin breaking parameters would lead to a substantial change in $Y_p$.  In one extreme, $\D M_{n-p} > \D M_{n-p}^\textrm{PDG}$, there would be very little primordial ${}^4$He as the neutrons would all decay before they could be captured in D and ultimately ${}^4$He.
In the other extreme, $\D M_{n-p} < \D M_{n-p}^\textrm{PDG}$, the neutron lifetime would become significantly longer than the time-scale for BBN, and thus a large fraction of the neutrons would become captured in D and then ${}^4$He, producing a H depleted Universe.

%
\section{Conclusions \label{sec:concl}}
In the work described in this article, we use results from state of the art lattice QCD calculations to provide, for the first time, a rigorous and quantitative connection between Big Bang Nucleosynthesis and the fundamental parameters of the Standard Model which control isospin breaking: the electromagnetic coupling, $\afs$ and the splitting between the down and up quark masses, $\d=\frac{1}{2}(m_d-m_u)$.
These results may help constrain prospective theories for physics beyond the Standard Model, particularly those which couple to isospin violation.  For example, theories of dark matter which couple differently to the proton and neutron~\cite{Hill:2014yxa}.

The ${}^4$He mass fraction tracks almost perfectly the nucleon mass splitting: simultaneous variations in $\afs$ and $\d$ which leave $M_n - M_p = \D M_{n-p}^\textrm{PDG}$, result in unobservably small changes to the primordial abundance of ${}^4$He.
In contrast, the abundance of D is also sensitive to the modified fusion rates which are sensitive to $\afs$ and not very sensitive to $\d$.
Therefore, a combined analysis of the modification to ${}^4$He and D produces very tight constraints on the allowed variation of $\afs$ and $\d$ while maintaining consistency with the observed abundances of D and ${}^4$He.
At the 3-sigma level, these variations are restricted to be $\lesssim0.8\%$.
However, a simultaneous variation of both isospin breaking parameters relaxes this constraint to $\lesssim1.25\%$, provided the variations are correlated.
We also verified that primordial variations in $\afs$ and $\d$ from their values today, within the region constrained by ${}^4$He and D, are not capable of resolving the $^7$Li puzzle.

We are now in a new era in which lattice QCD applied to problems in nuclear physics can have a quantitative impact on both our understanding of nuclear physics as well as the search for new physics from beyond the Standard Model.
For example, prior work has determined the nature~\cite{Aoki:2006we} and temperature~\cite{Aoki:2006br,Bazavov:2011nk}, $T\sim150$~MeV, of the QCD phase transition at which point the quarks and gluons hadronize to form protons and neutrons.
This work represents a quantitative connection between the quarks and the cosmos during the epoch of nucleosynthesis, when the Universe had expanded and cooled to a temperatures of $10 \gtrsim T \gtrsim 0.01$~MeV.
These are but a few examples in a growing set of applications of lattice QCD to nuclear physics.

\section{Acknowledgments}
We thank Tom Luu for useful conversations and initial help with the Fortran code.
We thank Wick Haxton for useful conversations and comments.
M. Heffernan thanks G. Vahala for the use of his computing resources.
The work of M. Heffernan was supported in part by the Natural Sciences and Engineering Research Council of Canada (NSERC).
The work of P. Banerjee was supported partially by the National Natural Science Foundation of China (NSFC) under Fund No.11533006.
The work of A. Walker-Loud was supported in part by LDRD funding from Lawrence Berkeley National Laboratory (LBNL) and by the U.S. Department of Energy (DOE) under Contract DE-AC02-05CH11231, under which the Regents of the University of California manage and operate LBNL, by the Office of Advanced Scientific Computing Research, Scientific Discovery through Advanced Computing (SciDAC) program under Award Number KB0301052,
and under the DOE Early Career Research Program, Office of Nuclear Physics under FWP NQCDAWL.
\bibliography{bbn}

\end{document}